# Systematic Detection of Energy Regressions in Software Version Histories

Jérôme Maquoi[a,*], François Bechet[a], Luís Cruz[b], Benoît Vanderose[a] and Xavier Devroey[a]

[a]NADI, University of Namur, rue de Bruxelles 61, 5000, Namur, Belgium
[b]Delft University of Technology, Postbus 5, 2600 AA, Delft, Netherlands

ABSTRACT

**Context**: Green software engineering is emerging as a crucial response to the rising energy impact of information technology, especially in continuous development. However, challenges remain in devising automated methods to efficiently identify energy regressions across commits. In particular, little effort has been put into automatically detecting regressions at the commit level by identifying statistically significant changes in energy consumption.
**Objectives**: This paper presents EnergyTrackr, an approach for detecting energy regressions across multiple commits.
**Methods**: We empirically evaluated EnergyTrackr on 1,932 commits from five Java projects, assessing its ability to identify significant energy changes. We also compared a straightforward linear algorithm that analyzes the entire commit history with a more efficient dichotomous search algorithm.
**Results**: The evaluation demonstrates the effectiveness of EnergyTrackr in identifying significant energy changes across commits. Our findings highlight the efficiency of the dichotomous algorithm in uncovering overall software energy trends, achieving results comparable to the linear approach while reducing resource consumption and processing time.
**Conclusion**: We expect EnergyTrackr to help developers accurately monitor energy regressions and improvements in their projects and optimize their source code to reduce software energy consumption.

## 1. Introduction

As computers become increasingly central to daily life, concerns about their energy use are growing. Freitag et al. (2021) estimate that the Information and Communication Technology (ICT) sector currently accounts for about 2.1% to 3.9% of global carbon emissions, a share that is expected to increase over time. Recently, the research community has dedicated significant effort to exploring various facets of software sustainability and green software engineering (Le Goaer and Hertout, 2023; Hao et al., 2013; Pang et al., 2016; Bonvoisin et al., 2024; Penzenstadler et al., 2012; Pereira et al., 2017; Poy et al., 2025; Simon et al., 2025). Despite these advancements, the field remains in its early development, with numerous challenges yet to be addressed. One such challenge involves identifying *energy regressions*, an increase in energy consumption between two software versions. Detecting these regressions is complex due to numerous factors. Yet, this step is mandatory to establish a strong empirical basis for identifying common defects causing unnecessarily high energy consumption in the software.

One primary challenge involves accurately measuring software energy consumption. Several methodologies exist, spanning different levels of granularity, from individual lines of code to entire applications. In this work, we focus on the application level, as it is more portable and introduces less overhead (and potential biases) in measurement. Since we seek to identify energy regressions, we can leverage commit changes to identify possible underlying code patterns that cause them. We rely on the software's test suite execution to measure its energy consumption. While larger software is expected to consume more energy, significant or abrupt changes may indicate an underlying energy-related issue within the codebase.

Danglot et al. (2024) introduced the concept of *energy regression testing*. It expands the scope of regression testing (i.e., ensuring that recent changes have not introduced new defects in pre-existing functionality) to monitor changes in software energy consumption, enabling detection of variations in energy usage resulting from source code modifications. Complementing their approach, we seek to provide an overview of energy consumption throughout the project's history, including potential energy regressions.

This paper builds on our initial tool paper on EnergyTrackr (Bechet et al., 2026), an approach that helps minimize energy consumption in software development. In a nutshell, EnergyTrackr considers the entire commit history (or an arbitrary portion of it), builds each commit individually, and measures its energy consumption based on test case executions. It provides statistical analysis support with configurable heuristics to identify energy regressions and perform triage, as well as graphical representations of the history to help developers spot problematic code patterns that may be causing energy regressions. A difference in energy consumption between two commits is marked as a regression if the difference in their energy consumption, measured via the execution of their test suites, is statistically significant.

*Corresponding author
jerome.maquoi@unamur.be (J. Maquoi); francois.bechet@tutanota.com (F. Bechet); l.cruz@tudelft.nl (L. Cruz); benoit.vanderose@unamur.be (B. Vanderose); xavier.devroey@unamur.be (X. Devroey)
ORCID(s): 0009-0006-3576-0757 (J. Maquoi); 0009-0008-4533-0552 (F. Bechet); 0000-0002-1615-355X (L. Cruz); 0000-0001-9752-0085 (B. Vanderose); 0000-0002-0831-7606 (X. Devroey)

A major constraint of the current *naive algorithm* used by EnergyTrackr is that it requires processing the entire commit history. Although this has the advantage of being simple and measuring every single commit, the biggest downside of this approach is the time required to process the entire commit range. To address this limitation, we extended EnergyTrackr to include a more efficient algorithm based on *dichotomous search*.

We design and perform an empirical evaluation of EnergyTrackr to answer the following research questions:

- **RQ1**: To what extent can an automated energy analysis pipeline reliably detect statistically significant energy regressions across commits in real-world Java projects, based on the execution of the test cases?
- **RQ2**: To what extent can the dichotomous search algorithm approximate the naive search algorithm to identify energy regressions across ranges of commits in real-world Java projects, based on the execution of the test cases?

Our analysis of five Java projects demonstrates that both algorithms effectively detect significant energy regressions across multiple commits in all projects. The data indicate a high rate of normally distributed results for almost all projects. Findings suggest that the dichotomous algorithm generally yields results that are statistically and practically comparable to those of the naive algorithm for most commits in each project, except for one. The dichotomous algorithm effectively captures most of the trend directions identified by the naive algorithm, omitting only a few trends. Our contribution includes the replication package containing the data produced during our evaluation (Maquoi et al., 2026).

The remainder of this paper is structured as follows: Section 2 presents the background and related work. The EnergyTrackr approach is presented in Section 3, with its implementation described in Section 4. Section 5 describes our evaluation protocol, with the results presented in Section 6. Sections 7 and 8 discuss the results, current limitations, and new directions for future work. And Section 9 concludes the paper.

## 2. Background and related work

### 2.1. Software energy consumption measurement

There are three main methods to assess the energy consumption of software Fahad et al. (2019): (i) System-level physical measurements using physical power meters. (ii) Measurements obtained through on-chip power sensors. (iii) Energy predictive models.

*Physical power meters* Physical power meters provide accurate measurements but lack hardware-level detail Fahad et al. (2019), making analysis complex. They are cumbersome, requiring purchase, calibration, and setup. Banerjee et al. (2014) automates energy hotspot detection in Android using a power meter, while Schubert et al. (2012) employed *Watts Up? Pro* to evaluate the *eprof* profiler. Hindle et al. (2014) introduced GreenMiner, which measures mobile device energy, streamlines testing, and produces detailed reports. Though stable, it depends on costly and custom hardware. Unlike these methods, EnergyTrackr operates without physical power meters and is applicable beyond mobile devices.

*On-chip power sensors* The second approach uses manufacturer-provided on-chip power sensors, such as Intel's Running Average Power Limit (RAPL), which provides CPU electricity consumption data to the operating system (OS) Raffin and Trystram (2025). RAPL is widely regarded as one of the most accurate tools for measuring processor energy consumption Spencer Desrochers et al. (2016); Khan et al. (2018), explaining tis prevalent use in CPU power research Liu et al. (2015); Jay et al. (2023); Schuler and Kotsis (2022). In software energy consumption, literature primarily focuses on CPU measurement Pathania et al. (2025), justifying RAPL's widespread adoption. For instance, Ournani et al. (2021b) employ RAPL to compare energy use across Java I/O libraries and methods. RAPL also integrates with tools like PowerAPI Fieni et al. (2024), TLPC-sensor Danglot et al. (2024) and the Linux `perf` utility Perf Wiki (2026).

PowerAPI provides real-time, machine learning-based estimates of an application's power consumption from performance metrics, auto-calibrating as needed Fieni et al. (2024). In contrast, EnergyTrackr operates without machine learning, focusing on tracking changes in energy consumption across different Git commits.

TLPC-sensor collects energy consumption and performance metrics to analyze their relationship Danglot et al. (2024). It uses RAPL for time-based measurements, controlled with explicit *start*() and *stop*() calls, enabling precise and granular measurements of specific code blocks. While flexible, this instrumentation adds complexity. In contrast, EnergyTrackr does not require modification of the source code for energy measurements.

`Perf` is a widely recognized Linux utility for collecting energy measurement data. One key advantage of utilizing `perf` over manual reading of RAPL driver files is its higher sampling rate and lower overhead, due to its integration within the Linux kernel. Due to its ability to deliver high-quality energy measurements, this study employed this tool to gather relevant data. In a previous study, the authors compared `perf` with alternative RAPL-based tools, concluding that these tools produce highly consistent results Jay et al. (2023). The findings indicate that `perf` is sufficiently reliable when complemented with additional techniques such as repeated measurements. EnergyTrackr uses `Perf` to measure the energy consumption associated with commits in Java projects.

*Energy predictive models* The third approach employs theoretical models to estimate consumption without direct measurements. For example, Acar et al. (2016) used formulas to approximate CPU, memory, and disk power during execution, avoiding extra infrastructure and enabling detailed analysis. However, their accuracy is often limited McCullough et al. (2011); O'brien et al. (2017); Shahid et al. (2017). In contrast, EnergyTrackr uses `perf` and the RAPL interface for precise CPU energy measurements.

## 2.2. Challenges of reliable energy measurement

*Granularity* Power measurement varies in granularity, from individual code lines to total device energy use. Harman et al. (2015) identify three levels: (i) Fine-grained level, which evaluates the contribution of each line of code to the overall energy consumption. For example, Li et al. (2013) describe an approach that provides developers with source line level energy information by combining hardware-based energy measurements with program analysis and statistical modeling techniques; Maquoi et al. (2025) identify energy-intensive source code constructs by analyzing the correlation between energy consumption and individual lines of Java code, and suggest a link between energy consumption and the number of objects' attributes created within that code. (ii) Mid-grained level focuses on the energy utilized by a specific block, method, or procedure within the program. For example, Hähnel et al. (2012) use RAPL to measure the power consumption of short-running paths, such as the energy cost of a single function within an application or a system call of the underlying OS; Schubert et al. (2012) present eprof, a software energy profiler that can identify the energy used by individual functions. EnergyTrackr lies within the mid-grained granularity level, as the objective is to find energy regressions across commits. (iii) Coarse-grained level measures the total energy consumption associated with executing the program or process over a designated period. For example, Mancebo et al. (2021) present the FEETINGS framework, designed to measure, analyze, and visualize the energy consumption of software. This framework investigates possible correlations between software maintainability metrics, such as the total lines of code, and the corresponding energy consumption.

*Repoducibility* Energy and power measurements in scientific studies are complex and influenced by factors like component and ambient temperatures, background processes, and thread scheduling (Cruz, 2021). This variability complicates assessing software energy efficiency, as single measurements can produce inconsistent results across trials.

Cruz (2021) proposed a method widely adopted in green software engineering (Roque et al., 2025; Xiao et al., 2025; Georgiou et al., 2022), and employed in developing EnergyTrackr:

**Zen mode**: Minimize the computer environment impact, for example, by closing as many background processes as possible or by turning off auto-brightness mode on the devices.

**Freeze your settings**: Save the current state of the machine and reuse that same setting during the entire testing process.

**Warm-up**: A CPU performs more efficiently when its temperature stabilizes after idle. A calibration or warm-up phase is essential before measurements to reach the target temperature, preventing fluctuations from affecting initial results and ensuring accuracy, as noted by Ournani et al. (2021a).

**Repeat**: To address measurement instability, perform multiple runs Arcuri and Briand (2014) and use the median as the final value. Cruz (2021) recommends 30 measurements, but practical constraints often lead to fewer, such as 10 repetitions (Bangash et al., 2017). Some studies, like Danglot et al. (2024), use up to 100 repetitions.

**Rest**: Temperature influences CPU energy efficiency, causing a tail effect where the end of one measurement affects the next. To mitigate this, Cruz (2021) recommends resting the machine between tests—typically around one minute—to allow temperatures to stabilize. While it does prolong the measurement process, this duration balances accuracy with evaluation cost.

**Shuffle**: Repeating measurements for the same commit can be affected by environmental factors like background processes. Creating batches of tasks, e.g., measuring $n$ commits $k$ times results in $n \times k$ tasks, can help mitigate this variability. Randomizing the order of these tasks can further reduce environmental impact and improve the reliability of measurement results.

**Keep it cool**: A CPU getting too hot can start throttling and impacting the measurements. Therefore, ensuring the CPU stays cool during the process is very important. Collecting CPU temperature data during the measurements can also be a good way to ensure it does not impact the measurements.

**Automate executions**: Reproducibility is impossible without automating the whole measurement process. Writing a script for the measurement execution is a mandatory step.

These methodological recommendations improve the reliability of energy measurements but increase the cost of experimentation, as each measurement must be repeated multiple times to account for variability.

## 2.3. Energy regressions

Energy regression testing is defined by Danglot et al. (2024) as detecting energy consumption drifts in software systems induced by changes observed at the source code level, a.k.a energy regressions. Their study emphasizes the challenge of effectively notifying developers about increased energy consumption in applications during CI processes. To do so, they gather energy measurements via the execution of the tests of the analyzed software at specific commits. Only the tests that cover the source code changes introduced by each commit are executed to optimize testing efficiency. The study suggests that developers' tests can identify energy regressions resulting from code changes, although the effectiveness of this approach varies depending on the configuration. This study is a foundation for further research, as additional work remains. Conversely, EnergyTrackr does not utilize Continuous Integration (CI) for energy regression detection, as RAPL often does not have access to such development environments. Instead, it employs `perf` to measure the energy consumption associated with code commits accurately. In addition to this methodology, EnergyTrackr offers a comprehensive overview of the project's energy consumption history, including detection of any potential regressions in energy efficiency.

The links between energy and runtime performance ahve been extensively studied (Pereira et al., 2017; Weber et al., 2023; Aragón-Jurado et al., 2025). Although they often correlate, they do not always evolve together. Pereira et al. (2017) found that although some languages are faster and more energy-efficient, faster languages are not always the most energy-efficient. Cruz and Abreu (2017) found that some mobile performance optimizations enhance energy efficiency, but others have no effect or worsen it, indicating no universal solution. Consequently, performance detection techniques cannot be directly applied to energy regressions. They also highlight the absence of mature tools to help developers identify energy hot spots and regressions.

The source code may be a cause of software energy regressions. Poy et al. (2025) conducted a systematic mapping study examining how design patterns, code smells, and refactoring techniques influence software energy consumption. Their findings indicate that the usage of design patterns does not always positively impact software energy consumption. For example, positive impacts are observed in some cases, such as with the Flyweight and Interpreter patterns. In contrast, other patterns, including the Chain of Responsibility, Decorator, or Iterator, were associated with increased energy consumption. The results were inconclusive for specific patterns like Abstract Factory, Adapter, and Bridge due to conflicting findings across primary studies.

Concerning the code smells, the topic is complex. Poy et al. (2025) suggests that removing nearly all code smells in their primary studies enhances software energy efficiency. However, Connolly Bree and Ó Cinnéide (2025) conducted a systematic literature review examining the impact of various design elements, including code smells, on software energy consumption. Their findings indicate that removing code smells does not always reduce energy usage. This may be because identifying and addressing code smells can be more challenging than implementing design patterns. Further research is needed to understand better instances where refactoring efforts may cause energy regressions related to code smells.

In contrast, EnergyTrackr does not analyze individual design elements or evaluate the effects of specific design patterns against unpatterned source code. Instead, it is tailored to identify energy regressions by studying the evolution of code commits over time and can serve as a basis for future research on those topics.

## 2.4. Energy consumption across software evolution

Mining Software Repositories (MSR) has become an established research field, and is intended to study software evolution through the analysis of version-control histories (Kagdi et al., 2007; de Freitas Farias et al., 2016).

Researchers have examined how non-functional properties like runtime performance evolve, focusing on identifying regressions between versions. Generally, benchmarks compare recent releases against earlier versions to detect performance regressions or improvements (Bezemer et al., 2014; Heger et al., 2013; Sandoval Alcocer et al., 2016). These methods show that analyzing the evolution of performance yields valuable insights beyond the isolated version assessments.

Despite growing interest, most longitudinal studies on software energy efficiency examine only a few versions. Bangash et al. (2017) linked energy consumption to software structure using 45 versions of three apps. Cruz et al. (2019) analyzed thousands of commits to assess the impact of energy-oriented changes on Android app maintainability, but did not focus on the general evolution of energy. Overall, longitudinal energy studies are limited, leaving developers and researchers without comprehensive insights into long-term energy trends or the identification of energy hot spots.

EnergyTrackr overcomes these limitations by enabling developers and researchers to automatically measure and analyze their software's energy consumption across hundreds of commits within Java projects, and to generate comprehensive reports that identify potential energy regressions and improvements observed within the analyzed commits.

## 2.5. Scalability of historical energy analysis

Energy measurements are noisy and require methodological precautions explained in the section 2.2. While these improve reliability, they greatly increase time and resource costs (Hindle, 2015). That means that the measurement cost grows rapidly with the number of versions analyzed. Historical energy analysis is therefore a scalability problem.

Detecting a regression and finding its origin are different tasks. Developers and researchers need to identify which commit introduced the regression to be able to find its cause. Search-based strategies can exploit the ordered nature of commits histories to reduce the number of measurements. For example, Sandoval Alcocer et al. (2016) proposed a sampling technique whose objective is to identify software versions that are prone to performance regressions without having to measure them all. A key challenge for such experimentation is to balance localization accuracy, measurement cost and execution time.

To address this scalability challenge, EnergyTrackr offers two different search algorithms: a straightforward approach and a dichotomous approach. The straightforward method provides precise localization of energy regressions, but requires increased resources and processing time. The dichotomous method identifies trends in the software's energy consumption, optimizing for resources and time efficiency, though with a potential trade-off in localization accuracy.

## 3. EnergyTrackr approach

In a nutshell, the core idea of EnergyTrackr is as follows: (i) running over a commit range of a project under test to measure its energy consumption in Joules and its runtime execution in seconds, based on the execution of the test cases; (ii) sorting the result, as EnergyTrackr runs tasks in a random order to counter environment effects on energy measurements; and (iii) building a report, implemented as a web page (allowing for easy sharing), containing the analysis of the results and flagging potential energy regression commits.

### 3.1. Commit range processing

EnergyTrackr can use two different detection algorithms to detect changes in energy consumption: a *naive algorithm* and a *dichotomous algorithm*. The naive algorithm measures each commit individually in the project under test and compares its energy consumption to that of the next commit. The dichotomous algorithm measures only a subset of the project's commits by splitting the history into fixed-size *commit regions*. Only the first and last commits of each region are measured. If a change in energy consumption is detected between those two commits, the region is split into two smaller regions and the algorithm loops until there are no more regions to analyze.

Figure 1 shows the high-level diagram of the algorithm. The algorithm starts by creating the initial regions and preparing the machine (as described in Section 3.2). This preparation is executed only once for all the measurements. To follow best practices and avoid threats to validity (as presented in Section 3.3), measurements are conducted on a set of commits (i.e., endpoints) and not on individual commits. Each measure is repeated 30 times per commit to account for possible randomness. Based on the collected data, *change points* are detected (as presented in Section 3.4), and each region analyzed is classified as *pruned* (i.e., no change was detected), *leaf change point* (i.e., a change was detected, but the region cannot be split anymore), or is split in half for further analysis. When no pending regions remain, the algorithm ends and EnergyTrackr produces a report (as explained in Section 3.5).

While the naive method has the advantage of being complete and measuring every commit, its biggest downside is the time required to process the entire commit range. The dichotomous algorithm addresses this limitation by following a divide-and-conquer strategy. The main idea is to split the project's history into multiple commit regions. The initial regions' size is calculated like this:

$$\text{Initial regions' size} = \max\left(5, \left\lceil \frac{n}{20} \right\rceil\right)$$

where n is the total number of commits in the commit range. This enables adaptation to the number of commits given as input, with a minimum of 5. For example, if the total number of commits is 500, the initial regions' size will be 25, so there will be 20 regions of 25 commits each. Small initial regions will detect more regressions but take longer. From our experience, this is a safe balance between regression detection and execution time. However, the initial size can be customized, depending on the use case. An interesting property of the algorithm is that if the initial region size is set to 2, it is equivalent to the naive algorithm, which tests all commits by comparing each commit to the next one.

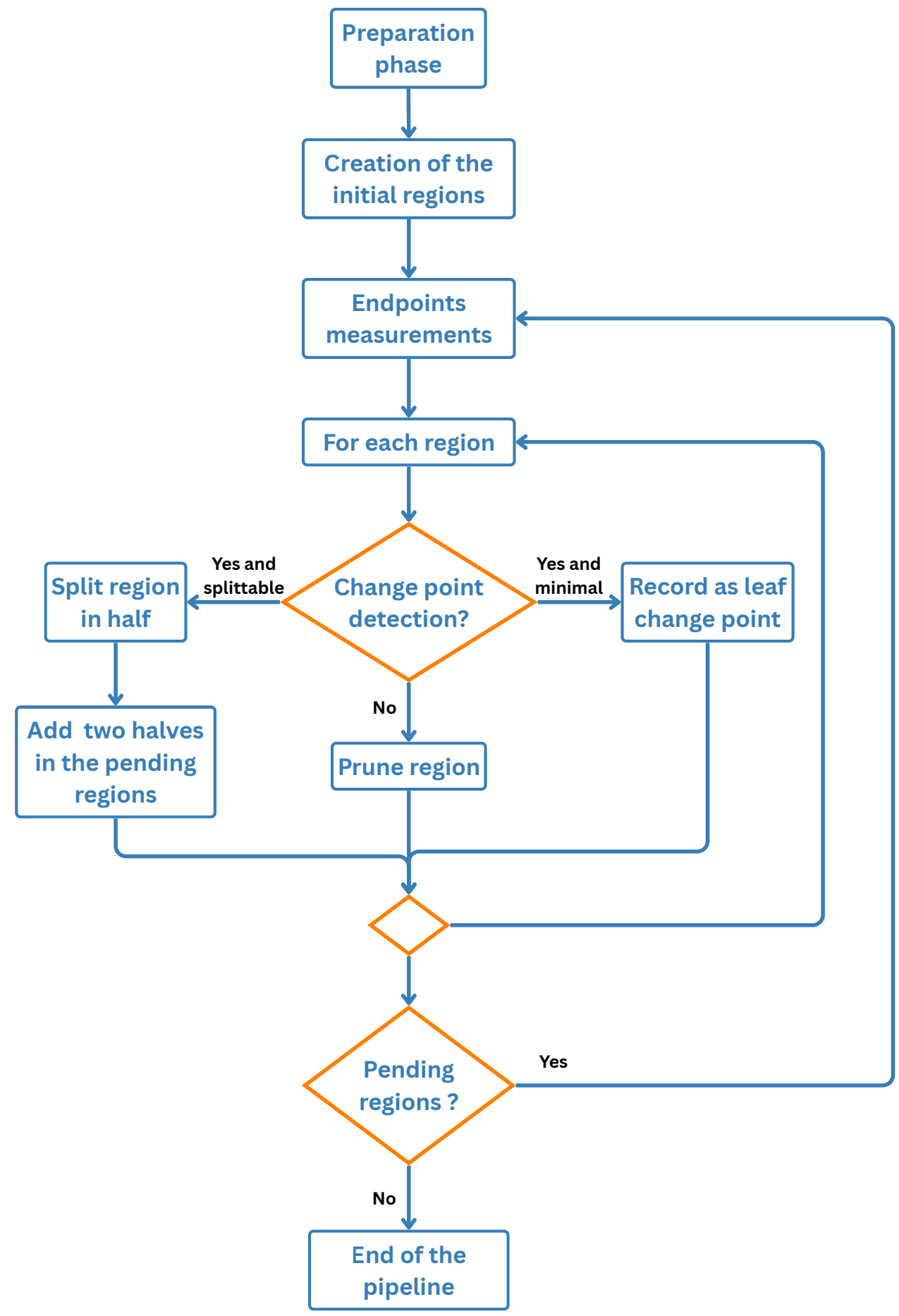

**Figure 1:** High-level pipeline of the dichotomous algorithm

Figure 2 shows an example of exploration of commits with the dichotomous algorithm. Let's assume we have n commits ($c_0 \rightarrow c_n$) of a project that we want to analyze. These commits are split into two initial regions (from iteration 0 to iteration 1): $c_0 \rightarrow c_k$ and $c_k \rightarrow c_n$). These two regions are measured and a change point was detected in both, so they are split in half from iteration 1 to iteration 2. $c_0 \rightarrow c_k$ is split into $c_0 \rightarrow c_j$ and $c_j \rightarrow c_k$, and $c_k \rightarrow c_n$ is split into $c_k \rightarrow c_m$ and $c_m \rightarrow c_n$. From iteration 2 to iteration 3, the four regions are measured, and the regions $c_j \rightarrow c_k$ and $c_k \rightarrow c_m$ do not have a change point, so they are pruned. The other regions are split in half again. The same process is done one final time from iteration 3 to iteration 4. At the end, only leaves ($c_0 \rightarrow c_h$ and $c_m \rightarrow c_i$) and pruned versions ($c_h \rightarrow c_j$, $c_i \rightarrow$ $c_l$ and $c_l \to c_n$) remain, so the pipeline stops. In the end, this pipeline identified 2 regressions or improvements.

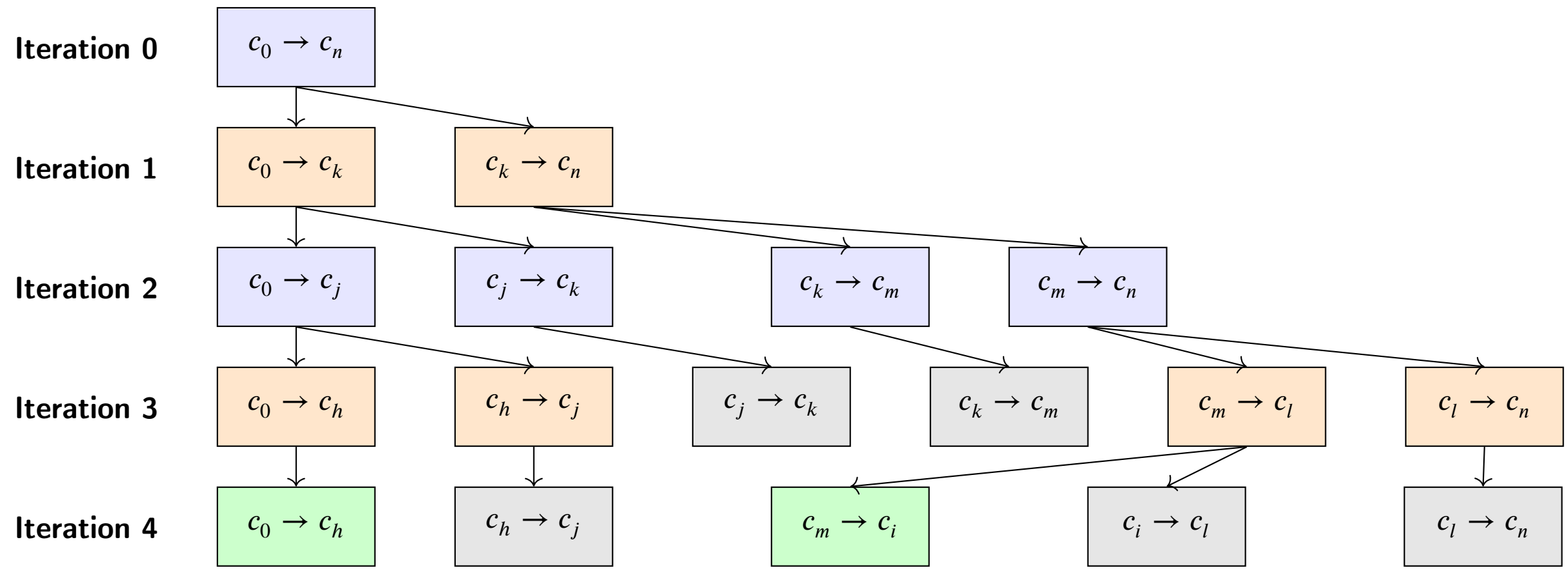

**Figure 2:** Example of exploration of commit regions across iterations in the dichotomous algorithm. Blue = unexplored region; Orange = split regions; Green = regression/improvement detected (leaf); Gray = pruned region.

## 3.2. Preparation phase

This phase stabilizes the Linux machine for repeatable and stable energy measurements by turning off power-saving features and pinning frequencies (**Zen mode**) via a configuration file. All modifications are reversible, as the device parameters are copied rather than directly altered, allowing users to easily switch between defined configurations of EnergyTrackr (**Freeze your settings**).

During this preparation phase, EnergyTrackr clones the target repository and collects the relevant commit history for analysis. An initial preprocessing stage is conducted, which involves filtering out unnecessary commits, e.g., commits that do not update any Java file. Our implementation also includes a check that `perf` can be used without root permissions.

*Stability verification* A stability verification step is incorporated before conducting the measurements (**Warm-up**). This step aims to confirm that the system is in a stable and consistent energy state. This step can also be executed in the following stages if the environment is unstable. During this process, the modified z-score (Iglewicz and Hoaglin, 1993) statistical method will be employed to identify anomalous data points from energy values (directly obtained from the RAPL interface in our implementation). The modified z-score $z$ for a sample $x$ is computed as:

$$z = \frac{0.6745 \cdot (x - \tilde{x})}{\text{MAD}}$$

where $\tilde{x}$ is the median of the samples, $\text{MAD} = \widetilde{\left(|x_i - \tilde{x}|\right)}$ is the Median Absolute Deviation, and $x_i$ are the individual sample values.

This stability verification process begins by collecting five initial warm-up samples to establish a baseline. Subsequently, measurements are taken from 30 additional samples, each individually compared to the established baseline. If any measurement's modified z-score exceeds the predefined threshold of 3.5, the system is flagged as potentially unstable, and the process is halted. The modified z-score is utilized instead of standard deviation due to its robustness in handling non-normal data distributions and its reliability with small or skewed datasets. Energy differences are analyzed rather than absolute energy values, as the latter can increase over time (e.g., the RAPL interface reporting cumulative energy consumption since system startup). In this context, the rate of change between samples is of primary interest for stability assessment.

## 3.3. Measurement phase

The selected approach for energy measurements involves wrapping the entire test suite execution (using `perf` in our implementation) rather than instrumenting individual tests to achieve greater precision. This method offers benefits such as reduced overhead and improved measurement accuracy. Specifically, the increased sampling duration enables more reliable statistical analysis by mitigating the impact of environmental variability, such as background processes.

During this phase, commits are organized into batches, which are randomized to counter environmental effects on measurement quality (**Shuffle**). After that, all commits within each batch are built. The energy consumption measurement is done based on the test suite execution. The test execution command can be configured to accommodate different build systems and test environments (e.g., `mvn test` for Java projects). The pipeline can be configured to run the same commit a specified number of times (**Repeat**), with the default setting of 30 repetitions as recommended by Arcuri and Briand (2014). In our implementation, a custom bash script was designed to set up the machine on which EnergyTrackr will run (**Automate executions**).

The current CPU temperature is monitored to ensure it remains within safe operational limits (**Keep it cool**). If temperatures are too high, the process pauses until cooling conditions are achieved (**Rest**). Energy consumption measurements, runtime execution and CPU temperature before and after the measurements are then collected for each commit in the batch and compiled for analysis.

## 3.4. Change point detection

Given the randomized execution of the measurements, the data needs to be sorted before analysis. In our implementation, this step produced CSV files. The data for the different commits are grouped by commit and sorted in descending order by date before being preprocessed.

### *3.4.1. Data preprocessing*

The first stage checks, for each commit, whether the data follows a normal distribution using the Shapiro-Wilk test (Shapiro and Wilk, 1965) for the different energy measures for that commit. It is important to note that commits with non-normal distributions can still be classified as significant changes. This is because distribution normality is not part of the significance classification itself but serves as an additional measurement quality indicator.

Second, following best practices (Cruz, 2021), we filter out outliers. Specifically, this second stage detects transient outlier commits, i.e., drops in sustained energy that indicate potential outliers rather than real regressions, by using the interquartile range (IQR) statistic, with a Tukey multiplier of 1.5 (Tukey, 1949). The stage accepts some configuration variables as inputs, such as the rolling window range of commits used to detect outliers, the Tukey multiplier for IQR filtering, the maximum number of transient outlier commits, and the aggregation method used to combine multiple measurements for each commit.

### *3.4.2. Change detection*

This third stage identifies significant changes or regressions in the energy data and sorts them into five levels. A commit is first tested at level 1, then at level 2, etc., until level 5. The highest level the change satisfies determines its classification, meaning a change can satisfy multiple levels or fail some intermediate ones. As those settings can be project-dependent, they can be fully configured. We report here the levels with the default settings used in our implementation (and evaluation):

**Level 1**: Performs a parametric Welch's unequal variances t-test (Welch, 1947) to assess statistical significance with a threshold $\alpha = 0.05$. If the commit passes this test, then it is classified as *level 1* and considered a regression/improvement.

**Level 2**: Calculates Cohen's $d$ (Cohen, 1988) to measure effect size. It uses the following thresholds to classify the change: negligible for $|d| \leq 0.2$; small for $0.2 < |d| \leq 0.5$; medium for $0.5 < |d| \leq 0.8$; large for $|d| > 0.8$. If the commit is at least in the small category ($0.2 < |d|$), then it is classified as *level 2*.

**Level 3**: Computes the relative change $\Delta\%$ on the median energy values. Let $m_b$ be the median of the baseline sample and $m_t$ be the median of the test sample, the relative change for a pair of commits is computed like:

$$\text{Percentage Change} = \left| \frac{m_t - m_b}{m_b} \right|$$

It uses the following thresholds to classify the change: minor for $\Delta\% < 5.0\%$; moderate for $5.0\% \leq \Delta\% < 10.0\%$; and major for $\Delta\% \geq 10.0\%$. If the commit is at least moderate, then it is classified as *level 3*.

**Level 4**: Computes the practical significance $\Delta J$, i.e., how strong the regression is relative to the baseline. Let $m_b$ be the median of the baseline sample and $m_t$ be the median of the test sample, the relative change for a pair of commits is computed like:

$$\Delta J = m_t - m_b$$

It uses the following thresholds to classify the change: info for $\Delta J < 5.0\% \times m_b$; warning for $5.0\% \times m_b \leq \Delta J < 10.0\% \times m_b$; critical for $\Delta J \geq 10.0\% \times m_b$. If the commit is at least in the warning category, then it is classified as *level 4*. Improvements will consistently be categorized under the "info" classification and, as a result, will not attain level 4. This is because they will always exhibit a negative $\Delta J$ and consequently remain below 5.0% of $m_b$.

**Level 5**: When context tags from EnergyTrackr's configuration are present in the commit message, then the commit will be flagged to *level 5*. For example, if the commit message contains the word '*performance*' and that the tag '*performance*' is in the tag list.

Cruz (2021) and Danglot et al. (2024) inspired the design of levels 1 to 3. The additional levels were designed to provide a more exhaustive and comprehensive classification of changes and were developed specifically for EnergyTrackr. This combined approach ensures that only relevant, meaningful changes are reported, filtering out random fluctuations while highlighting regressions or improvements developers should address in priority.

## 3.5. Data visualization and report generation

To enable developers to analyze the results and identify energy regressions based on our classification from levels 1 to 5, EnergyTrackr proposes different energy consumption data visualizations.

### *3.5.1. Data plots*

This stage creates interactive plots to enable data visualization. The implementation of EnergyTrackr supports the following plots:

**Evolution plot** plot shows the history of the commit range. For instance, Figure 3 presents the evolution of energy consumption for a portion of Simple-OpenAI's commit history. Each blue circle indicates a commit. The red and green vertical bars represent commits with a level of at least 2, whether they signify an energy regression (red) or an improvement (green). Some commits are highlighted with a colored circle. These circles denote the level of each commit, starting from level 2, with size and color corresponding to the specific level: a blue circle for level 2, an orange circle for level 3, a purple circle for level 4, and a red circle for level 5. For example, to the right of Figure 3, a level

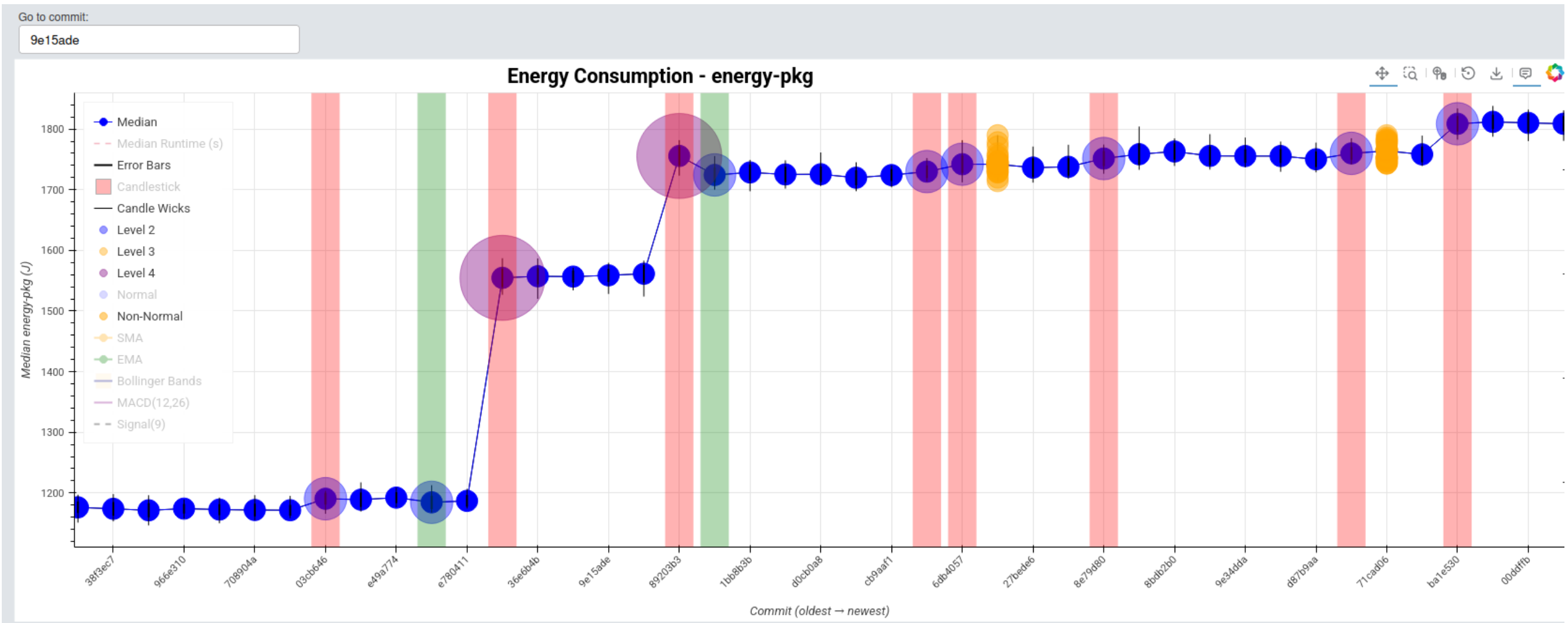

**Figure 3:** Evolution plot showing the commit `9e15ade` and nearby commits from `Simple-OpenAI` by using the naive algorithm.

4 regression is visible, indicating that this commit has been classified as "warning," with a negative impact estimated between 5.0% and 10.0% compared to the previous commit.

**Boxplots and violin plots** allow for comparing two commits in terms of the energy consumption distribution across the repeated executions of their respective test suites (Hintze and Nelson, 1998; Dutoit, 2012).

**Quantile-Quantile plot** compares the distributions of the energy consumption of two commits (Wilk and Gnanadesikan, 1968).

**Bootstrap plot** is an interactive histogram that helps confirm if the difference between two commits is statistically significant (Efron and Tibshirani, 1993).

**Cusum plot** shows the cumulative sum of the energy of improvements and regressions. It gives an overview of the macro scale of a project to assess the global energy impact.

**Change point detection plot** detects change points in the median energy data. It gives another overview of a project's abrupt changes in energy consumption.

### *3.5.2. Report generation*

The final stage of EnergyTrackr is to gather the different plots and generate a report with additional information, such as the level's configuration with the various thresholds, and a table listing the different commits with their metadata and statistics. This report can serve as a basis for developers to identify and debug energy regressions.

## 4. Open source implementation

EnergyTrackr is implemented as an open-source tool (Bechet et al., 2026).[1] It offers a pipes-and-filters design pattern that provides great modularity. Each stage of the tool is implemented as a separate module, enabling easier testing and future extensibility. It is possible to add a new stage to EnergyTrackr by defining a new module, or to modify the current behavior by removing or altering a stage in the configuration file. This choice of modular design with proper separation between configuration, energy measurement, plotting, and shared utilities was deliberate to ensure: (i) cross-language capabilities in both the analysis of different language projects (currently, only Java projects are supported), but also allowing interaction with other measurement tools written in different languages through an easier API interaction; (ii) better reproducibility thanks to a centralized configuration file, which is an important quality factor when taking energy measurements; and (iii) easier integration into existing developers' workflow.

The implementation includes two main configuration files to configure EnergyTrackr for projects running in different environments. The initial configuration file enables developers to define variables for the measurement pipeline, such as the repository URL, the number of commits to iterate over, and the granularity level (e.g., commit, branch, or tag). The second configuration file allows users to define which data transformations to apply and how to configure them, specify which plots to generate and their settings, and determine which page components to include and customize to build the final report.

## 5. Evaluation setup

The evaluation aims to validate our approach to energy regression detection (**RQ1**) and compare the naive and dichotomous search algorithms for energy regression detection (**RQ2**). This evaluation is quantitative and involves executing EnergyTrackr with both algorithms across a selection of real-world GitHub projects with strong test suites, measuring energy consumption, performing the data analysis, and building the reports (**RQ1**). The data of both algorithms will be compared in terms of energy regression identification and execution time (**RQ2**).

[1]Available at `https://github.com/snail-unamur/energytrackr/releases/tag/1.0.0`

**Table 1**
List of selected projects with their characteristics.

| | maker | fastexcel | classgraph | simple-openai | modelmapper |
|---|---|---|---|---|---|
| **Commit SHA** | b220def | b616428 | a03ed56 | 7af0d7d | 14145e1 |
| **JDK version** | 17 | 1.8 | 1.8 | 11 | 1.8 |
| **Lines of Code** | 5,956 | 2,222 | 12,291 | 3,814 | 3,449 |
| **Tot./Failed/Ignored** | 437 / 0 / 0 | 81 / 0 / 0 | 217 / 2 / 3 | 217 / 0 / 1 | 613 / 0 / 5 |
| **Line Coverage** | 98% (5854 / 5956) | 87% (1950 / 2222) | 65% (8072 / 12291) | 42% (1622 / 3814) | 85% (2952 / 3449) |
| **Branch Coverage** | 92% (3101 / 3449) | 70% (699 / 997) | 55% (5297 / 9493) | 41% (148 / 358) | 81% (1572 / 1918) |

To answer our RQs, we selected five Java projects, based on the following inclusion criteria: (i) a *long commit history* with at least 500 commits to allow a rich comparison; (ii) a *good test suite* that can be executed reliably over the commit history; (iii) *open source* projects recommended in the *Awesome-Java* list of open source projects;[2] (iv) the projects should be *maintained*; (v) relying on the *dependency management system Maven*, which is important to be able to run tests over a long history; and (vi) the project should use the *Git version control system*.

The selected projects are `maker`, `fastexcel`, `classgraph`, `simple-openai`, and `modelmapper`, listed in Table 1 with their characteristics, derived from the most recent commit used for each project. `Maker` is a lightweight Java class generator library. `Fastexcel` is designed to read and write big Excel files quickly. `Simple-openai` is a Java library designed to use the OpenAI Api. `Modelmapper` is an object mapping library that automatically maps objects to each other.

All the experiments were executed on a dedicated Ubuntu 22.04.5 server with 64 Intel(R) Xeon(R) Gold 6326 (2.90GHz) and 256GB of RAM. As explained in section 3.2, the CPU frequencies were pinned to 2.90GHz.

# 6. Evaluation Results

The evaluation, as well as all the data and analysis discussed in this section, are present in the replication package (Maquoi et al., 2026).

## 6.1. Energy regression detection (RQ1)

Tables 2 and 3 contain the results summary for each project, respectively, for the naive and dichotomous algorithms.

*Maker* Of the 320 commits analyzed with the naive algorithm for the `Maker` project, 88.75% of them passed the Shapiro-Wilk normality test: 114 significant changes were detected, and 68 were classified as regressions. Of the 34 commits analyzed with the dichotomous algorithm for the same project, 94.12% passed the Shapiro-Wilk normality test: 18 significant changes were detected, including 15 regressions.

The energy values range from 1396.33 to 1704.24 joules, with a range of 307.92 joules for the naive algorithm. For the dichotomous algorithm, the values range from 1393.95 to 1702.54 joules, with a range of 308.59 joules. This project consumes the least energy, with an average of 1587.95 joules for the naive algorithm and 1571.85 joules for the dichotomous algorithm. This range is shown in Figure 4(a).

*Fastexcel* On the 395 commits analyzed with the naive algorithm, 75.31% of them passed the Shapiro-Wilk normality test: 46 significant changes were detected, and 32 were classified as regressions. Of the 50 commits analyzed with the dichotomous algorithm for this project, 68% passed the Shapiro-Wilk normality test: 22 significant changes were detected, including 19 regressions.

The energy values range from 4227.04 to 8534.86 joules, with a range of 4307.82 joules for the naive algorithm. For the dichotomous algorithm, the values range from 4258.98 to 8500.47 joules, with a range of 4241.49 joules. This is the project with the widest energy range and the highest energy consumption, averaging 5913.17 joules for the naive algorithm and 5812.02 joules for the dichotomous algorithm.

*Classgraph* Of the 395 commits analyzed with the naive algorithm for the `Classgraph` project, 89.37% of them passed the Shapiro-Wilk normality test: 40 significant changes were detected, and 21 were classified as regressions. For the dichotomous algorithm, 32 commits were analyzed for the same project, and 90.63% passed the Shapiro-Wilk normality test: 13 significant changes were detected, and 10 were classified as regressions.

The energy values of `Classgraph` range from 2797.17 to 3216.85 joules for the naive algorithm, with a range of 419.67 joules, for the naive algorithm. For the dichotomous algorithm, the values range from 2805.55 to 3198.63 joules, with a range of 393.08 joules.

*Simple-OpenAI* Of the 386 commits analyzed with the naive algorithm for the `Simple-OpenAI` project, 84.46% passed the Shapiro-Wilk normality test: 65 significant changes were detected, and 40 were classified as regressions. Of the 48 commits analyzed with the dichotomous algorithm for the same project, 79.17% passed the Shapiro-Wilk normality test: 28 significant changes were detected, and 19 were classified as regressions.

The energy values range from 1167.82 to 4300.48 joules for the naive algorithm, with a range of 3132.66. For the dichotomous algorithm, the values range from 1020.64 to 3969.98 joules, a difference of 2949.34 joules.

[2] Available at `https://github.com/akullpp/awesome-java`

**Table 2**
Results naive algorithm summary

| | maker | fastexcel | classgraph | simple-openai | modelmapper |
|---|---|---|---|---|---|
| **Total commits** | 320 | 405 | 395 | 386 | 426 |
| **Commit range** | 2b9cfcc — b220def | f380749 — b616428 | 968401a — a03ed56 | a8d5a91 — 7af0d7d | 53e8482 — 14145e1 |
| **Commit dates** | (2021-05-28) — (2026-06-29) | (2022-09-21) — (2026-05-19) | (2022-03-02) — (2026-06-05) | (2023-08-24) — (2025-09-17) | (2023-07-19) — (2025-11-11) |
| **Significant changes** | 114 (68 regressions) | 46 (32 regressions) | 40 (21 regressions) | 65 (40 regressions) | 76 (53 regressions) |
| **Mean/Med./$\sigma$ energy (Joules)** | 1587.95 / 1618.43 / 90.99 | 5913.17 / 5912.99 / 910.14 | 2934.7 / 2918.98 / 99.49 | 2646.34 / 1818.41 / 1080.98 | 2599.11 / 2368.08 / 833.93 |
| **Mean/Med./$\sigma$ runtime (Seconds)** | 6.33 / 6.44 / 0.35 | 24.59 / 24.34 / 3.84 | 11.77 / 11.69 / 0.4 | 10.96 / 7.37 / 4.64 | 10.78 / 9.9 / 3.38 |
| **Normal dist.** | 284 / 320 (88.75%) | 305 / 405 (75.31%) | 353 / 395 (89.37%) | 326 / 386 (84.46%) | 307 / 426 (72.07%) |
| **Outliers removed** | 1 | 6 | 7 | 3 | 2 |
| **Rel. change ($\Delta\%$)** | 318 minor, 1 moderate, 0 major | 396 minor, 5 moderate, 3 major | 392 minor, 2 moderate, 0 major | 377 minor, 4 moderate, 4 major | 414 minor, 8 moderate, 3 major |
| **Pract. signif. ($\Delta J$)** | 318 info, 1 warning, 0 critical | 398 info, 3 warning, 3 critical | 393 info, 1 warning, 0 critical | 380 info, 2 warning, 3 critical | 415 info, 7 warning, 3 critical |
| **Level 2 / 3 / 4 / 5** | 109 / 1 / 1 / 3 | 33 / 7 / 6 / 0 | 33 / 6 / 1 / 0 | 56 / 4 / 5 / 0 | 61 / 3 / 10 / 2 |
| **Execution time** | 18 hours | 3 days, 16 hours | 1 day, 18 hours | 1 day, 15 hours | 1 day, 17 hours |

**Table 3**
Results dichotomous algorithm summary

| | maker | fastexcel | classgraph | simple-openai | modelmapper |
|---|---|---|---|---|---|
| **Total commits** | 34 | 50 | 32 | 48 | 66 |
| **Commit range** | 2b9cfcc — b220def | f380749 — f380749 | 968401a — 052eaf0 | 7a4785c — 7af0d7d | 53e8482 — 14145e1 |
| **Commit dates** | (2021-05-28) — (2026-06-29) | (2022-09-21) — (2026-05-19) | (2022-03-02) — (2026-06-05) | (2023-08-24) — (2025-09-17) | (2023-07-19) — (2025-11-11) |
| **Significant changes** | 18 (15 regressions) | 22 (19 regressions) | 13 (10 regressions) | 28 (19 regressions) | 33 (27 regressions) |
| **Mean/Med./$\sigma$ energy (Joules)** | 1571.85 / 1607.46 / 97.57 | 5812.02 / 5610.3 / 1092.19 | 2948.53 / 2947.14 / 100.22 | 2178.15 / 1616.14 / 1014.66 | 2645.61 / 2433.97 / 824.98 |
| **Mean/Med./$\sigma$ runtime (Seconds)** | 6.27 / 6.41 / 0.37 | 24.18 / 23.36 / 4.58 | 11.83 / 11.83 / 0.4 | 8.98 / 6.56 / 4.32 | 10.97 / 10.15 / 3.35 |
| **Normal dist.** | 32 / 34 (94.12%) | 34 / 50 (68%) | 29 / 32 (90.63%) | 38 / 48 (79.17%) | 52 / 66 (78.79%) |
| **Outliers removed** | 0 | 0 | 1 | 0 | 0 |
| **Rel. change ($\Delta\%$)** | 32 minor, 1 moderate, 0 major | 44 minor, 3 moderate, 2 major | 31 minor, 0 moderate, 0 major | 42 minor, 1 moderate, 4 major | 56 minor, 6 moderate, 3 major |
| **Pract. signif. ($\Delta J$)** | 32 info, 1 warning, 0 critical | 45 info, 2 warning, 2 critical | 31 info, 0 warning, 0 critical | 44 info, 0 warning, 3 critical | 56 info, 6 warning, 3 critical |
| **Level 2 / 3 / 4 / 5** | 15 / 1 / 1 / 1 | 13 / 5 / 4 / 0 | 12 / 1 / 0 / 0 | 21 / 4 / 3 / 0 | 22 / 2 / 9 / 0 |
| **Execution time** | 2 hours | 10 hours 30 minutes | 3 hours 30 minutes | 4 hours | 6 hours |

*Modelmapper* Of the 426 commits analyzed with the naive algorithm for the `Modelmapper` project, 72.07% passed the Shapiro-Wilk normality test: 76 significant changes were detected, and 53 were classified as regressions. For the dichotomous algorithm, 66 commits were analyzed for the same project, and 78.79% passed the Shapiro-Wilk normality test: 33 significant changes were detected, and 27 were classified as regressions.

The energy values range from 1539.81 to 4213.24 joules for the naive algorithm, with a range of 2673.43 joules. For the dichotomous algorithm, the values range from 1545.56 to 4166.66 joules, with a range of 2621.1 joules.

> **Answer to RQ1**: Over all the analyzed projects, almost all projects have a rate of normal distributions higher than 75% for the naive algorithm and higher than 79% for the dichotomous algorithm. The project `Modelmapper` has a lower rate of normal distribution, with 72.07% for the naive algorithm and 78.79% for the dichotomous algorithm, and `fastexcel` has a lower rate only for the dichotomous algorithm, with 68% of normal distribution.

> In summary, the EnergyTrackr approach enables the production of stable and consistent energy values across repeated test executions, as demonstrated by the rates of normal distributions. The data and report can serve as a strong basis for identifying energy regression.

## 6.2. Comparison between the naive and the dichotomous algorithms (RQ2)

Figure 4 presents the energy evolution across all commits for each project using both algorithms. `SimpleOpenAI` exhibits the largest difference, with mean energy consumption differing by 468.19 joules between naive and dichotomous algorithms. The other projects show smaller differences: `Maker` (16.1 J), `Fastexcel` (101.15 J), `Classgraph` (13.83 J), and `Modelmapper` (46.5 J). The commits analyzed by the dichotomous algorithm form a subset of those analyzed by the naive algorithm, enabling direct comparison. Measurements consistency is addressed in Section 6.2.1, with trend comparisons in section 6.2.2.

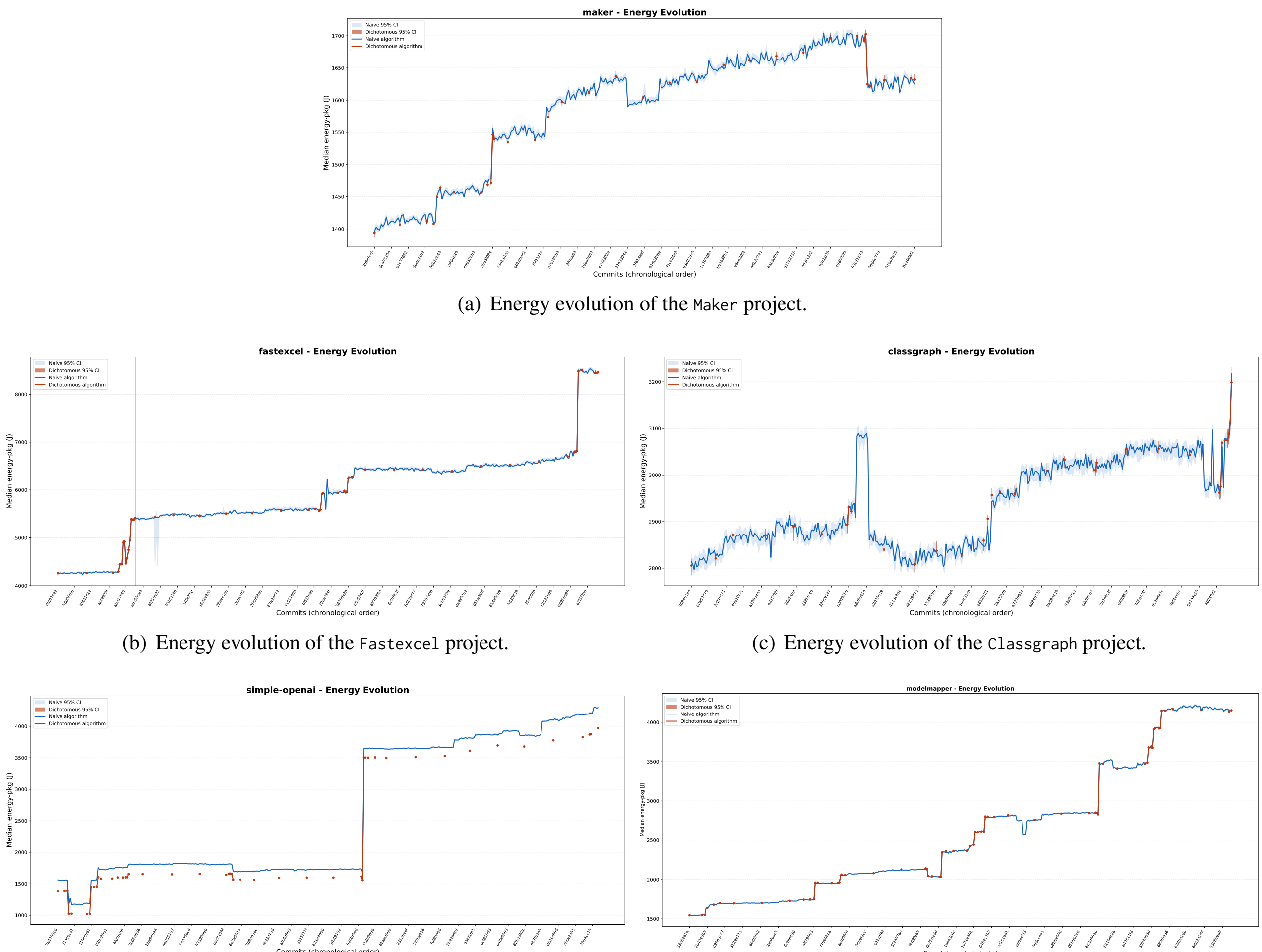

(a) Energy evolution of the `Maker` project.

(b) Energy evolution of the `Fastexcel` project.

(c) Energy evolution of the `Classgraph` project.

(d) Energy evolution of the `Simple-OpenAI` project.

(e) Energy evolution of the `Modelmapper` project.

**Figure 4:** Energy evolution across the entire range of commits for each project, using both the naive (blue) and dichotomous (orange) algorithms. An orange vertical line indicates a commit that was analyzed using the dichotomous algorithm but not by the naive algorithm.

### 6.2.1. *Consistency of naive and dichotomous measurements*

To evaluate the consistency of measurements between the two algorithms, the analysis focuses on commits present in both datasets, referred to as *shared commits*. The consistency of these shared commits was analyzed using three statistical methods: the nonparametric Wilcoxon rank-sum test to evaluate the null hypothesis that these commit samples originate from the same distribution, the Vargha-Delaney effect size to measure the magnitude of differences between naive and dichotomous commits (Arcuri and Briand, 2014), and the Kolmogorov-Smirnov test (KS) which evaluates if the two samples come from the same distribution or if they differ in shape (Massey Jr., 1951). The thresholds used for the Vargha-Delaney effect size follow those proposed by Vargha and Delaney (2000): $|\hat{A}_{12} - 0.5| < 0.14$ is classified as small, $0.14 \leq |\hat{A}_{12} - 0.5| < 0.21$ is classified as medium, and $|\hat{A}_{12} - 0.5| \geq 0.21$ is classified as large.

*Maker* Among the 34 shared commits analyzed, the Wilcoxon rank-sum test identified 3 commits (8.8%) with a statistically significant difference (p-value $< 0.05$), and the KS test identified 1 commit (2.9%) with distributional differences. Regarding effect sizes, 10 commits (29.4%) have a small effect size, and 3 commits (8.8%) have a medium effect size. These results indicate that the naive and dichotomous algorithms yield statistically and practically comparable results for the majority of commits.

*Fastexcel* Among the 50 commits analyzed by the dichotomous algorithm, one, abc42ae3, was not shared with the commits analyzed by the naive algorithm. This commit is represented by a vertical orange bar in Figure 4(b). Among the 49 shared commits analyzed, the Wilcoxon rank-sum test identified 2 commits (4.1%) with a statistically significant difference (p-value $< 0.05$), and the KS test identified 0 commits with distributional differences. Regarding effect size magnitude, 24 commits (49%) are classified as small, and 2 commits (4.1%) as medium. These findings suggest that the naive and dichotomous algorithms generate data that are statistically and practically comparable across the majority of commits.

*Classgraph* Among the 33 shared commits analyzed, the Wilcoxon rank-sum test pointed to 0 commits with statistically significant differences (p-value $< 0.05$). The KS test identified 0 commits with distributional differences. 9 commits (27.3%) were identified as having a small effect size. These results indicate that both naive and dichotomous algorithms produce statistically and practically compatible data for the majority of commits.

*Simple-OpenAI* The Wilcoxon rank-sum test identified all 48 shared commits as statistically different (p-value $< 0.05$). The same applies to the KS test, where 48 commits were identified as having distributional differences. The Vargha-Delaney test classified all 48 commits as having a large effect size. These results indicate that the naive and dichotomous algorithms yield statistically different measurements. The results of this test also indicate that all 48 commits of the dichotomous algorithm consume less energy than the shared commits for the naive algorithm. This is true because, for each commit, $\hat{A}_{12} > 0.5$. This confirms what we can see in Figure 4(d): all commits of the dichotomous algorithm consume less energy than the corresponding commits of the naive algorithm.

*Modelmapper* For this project, the results are mixed. Of the 66 shared commits, the Wilcoxon sum-rank test identified 17 of them (25.8%) with statistically significant differences, and the KS test identified 12 commits (18.2%) with distributional differences. Regarding magnitude, 25 commits (37.9%) were classified by the Vargha-Delaney test as small, 11 commits (16.7%) as medium, and 7 commits (10.6%) as large. For 21 commits (31.8%), $\hat{A}_{12} > 0.5$; for 1 commit (1.5%), $\hat{A}_{12} = 0.5$; and for 44 commits (66.7%), $\hat{A}_{12} < 0.5$. All of these results indicate that individual commits show disagreement, and that the majority of commits in the dichotomous algorithm tend to consume more than those in the naive algorithm.

#### 6.2.2. Comparison of trends

To compare the evolutionary trends between the naive and dichotomous algorithms, we use two metrics: the Spearman correlation (Spearman, 1904) and Trend Direction Accuracy (TDA). The Spearman correlation assesses whether the relative order of commits is maintained between the two algorithms, regardless of their absolute values. The TDA is inspired by directional accuracy defined by Pesaran and Timmermann (1992), which evaluates consecutive commits shared by both algorithms to determine whether the directions of their variations align. This is achieved by using the 95% confidence intervals of each commit to categorize each transition into one of three categories: significant increase, significant decrease, or stable. The common transitions between the two algorithms are then compared and the proportion of matching transitions is calculated as the TDA. A TDA approaching 100% indicates that the dichotomous curve accurately reflects the same trend directions as the naive curve.

*Maker* The Spearman correlation test indicates a strong preservation of the ranking between both algorithms ($\rho = 0.993$, $p$-value $= 6.099\text{e}{-31}$). Regarding the directionality of variations, the TDA analysis indicates that for the naive algorithm there were 15 upward, 3 downward, and 15 stable directions, whereas for the dichotomous algorithm there were 13 upward, 2 downward, and 18 stable directions. These results reveal that 84.8% of the variations (28 out of 33) were aligned in the same direction, as seen in Figure 4(a). The discrepancies observed may represent direct contradictions, such as one algorithm indicating an upward trend while the other indicates a downward trend, or differences in sensitivity, where one algorithm detects an upward change while the other remains stable. In the case of `Maker`, all 5 mismatches were attributable solely to differences in sensitivity.

*Fastexcel* The Spearman correlation analysis indicates a high degree of consistency in the rankings produced by both algorithms ($\rho = 0.995$, $p$-value $= 1.489\text{e}{-49}$). Regarding the directions of the variations, the TDA analysis indicates that for the naive algorithm there were 17 upward, 2 downward, and 29 stable directions, whereas for the dichotomous algorithm there were 16 upward, 1 downward, and 31 stable directions. These results reveal that 91.7% of the variations (44 out of 48) were aligned in the same direction, as seen in Figure 4(b). The 4 mismatches were attributable solely to differences in sensitivity.

*Classgraph* For this project, the Spearman correlation test indicates a strong preservation of rankings between the two algorithms ($\rho = 0.996$, $p$-value $= 1.18\text{e}{-34}$). Regarding the directions of the variations, the TDA analysis indicates that for the naive algorithm there were 10 upward, 3 downward, and 19 stable directions, whereas for the dichotomous algorithm there were 9 upward, 3 downward, and 20 stable directions. These results reveal that 96.9% of the variations (31 out of 32) were aligned in the same direction, as seen in Figure 4(c). The only mismatch was due to a difference in sensitivity.

*Simple-OpenAI* The Spearman correlation test indicates a strong preservation of the ranking between both algorithms ($\rho = 0.986$, $p$-value $= 2.331\text{e}{-37}$). Regarding the directions of variation, the TDA analysis indicates that for the naive algorithm there were 18 upward, 6 downward, and 23 stable directions, whereas for the dichotomous algorithm there were 16 upward, 8 downward, and 23 stable directions. The results reveal that 78.7% of the variations (37 out of 47) were aligned in the same direction, as seen in Figure 4(d). The 10 mismatches were attributable solely to differences in sensitivity.

*Modelmapper* The Spearman correlation test indicates a strong preservation of the ranking between both algorithms ($\rho = 0.998$, $p$-value = 8.18e−77). Regarding the directions of the variations, the TDA analysis indicates that for the naive algorithm there were 26 upward, 4 downward, and 35 stable directions, while for the dichotomous algorithm there were 23 upward, 6 downward, and 36 stable directions. These results reveal that 86.2% of the variations (56 out of 65) were aligned in the same direction, as seen in Figure 4(e). The 9 mismatches were attributable solely to differences in sensitivity.

> **Answer to RQ2**: Across all projects analyzed, `Maker`, `Fastexcel`, `Classgraph` and `Simple-OpenAI` demonstrate statistically and practically comparable results for the majority of commits with respect to both algorithms. This is supported by the outcomes of the Wilcoxon rank-sum test, the Kolmogorov-Smirnov test, and the Vargha-Delaney effect size. Specifically, for `Simple-OpenAI`, these tests indicate that the two algorithms produce significantly different measurements. However, all commit data generated by the dichotomous algorithm consume less energy compared to those from the naive algorithm.
> Regarding trend comparison, the Spearman correlation test applied across all projects shows that the relative ranking between the two algorithms remains strongly preserved within each project. The TDA analysis conducted on all projects reveals that the alignment of variation directions ranges from 78.7% to 96.9%. These findings suggest that the dichotomous algorithm captures most of the variation directions observed in the naive algorithm, with only a few trends omitted. Further detailed analysis indicates that any discrepancies across projects are due to differences in sensitivity rather than fundamental contradictions in the trends.

# 7. Threats to validity

## 7.1. Internal validity

*Energy measurement* Energy measurements are subject to many sources of variability, despite precautions like repeated measurements, randomized test execution, pre-setup of the computer for stability, and pre-measurement stability checks. Some measurement noise may remain due to measurement errors or hardware variability, like room temperature fluctuations, which was not controlled during this experiment. However, the machine is in a temperature-controlled server room, and thermal control was implemented within the measurement pipeline to avoid thermal throttling.

Batch size could influence results: smaller batches may weaken randomization, while larger batches delay measurements and increase variance from environmental noise. This factor was not explicitly tested and warrants future investigation.

Regarding ENERGYTRACKR's implementation, we took the following precautions: it includes unit tests (the project aims for 80% to 100% line coverage for the measurement modules). We also used linters (Ruff, Pylint, and Pyright) to identify potential defects.

*Changes in the test suite* Building on Danglot et al. (2024), we assume running tests with good coverage can detect energy regressions. While test-based measurements may be biased by test modifications, they obviate the need for manual custom tests, enhancing scalability without extensive expertise. We excluded commits with failing tests; including them could conflate test failures with true regressions. An alternative would be using mutation testing (Papadakis et al., 2019) to inject known regressions for precise evaluation, but we opted to focus on open-source projects to ensure broader relevance.

## 7.2. External validity

The evaluated projects varied in domain, but the small sample may limit generalizability. We selected Git-based open-source projects from the *Awesome-Java* list, focusing on popularity and ensuring reproducibility, with at least 500 commits. While prioritizing high code coverage, finding multiple projects meeting both commit activity and coverage criteria was difficult, thus we also included projects like `Simple-OpenAI` with 42% of line coverage and 41% of branch coverage.

This selection is based on practical considerations, but a larger, more diverse set of benchmarks is needed for better generalizability. The chosen projects, centered on Maven dependency management, may not represent the typical characteristics of an average Git repository.

## 7.3. Construct validity

Using RAPL for energy measurements is limited to overall CPU consumption, offering no detailed insight into specific sources like memory, storage, or network devices that may cause energy increases. While the CPU often dominates device's power use (Vasques et al., 2019), other components such as the power supply unit, motherboard, peripherals, memory, disks, and fans also consume power and are not accounted for by RAPL.

Thermal power generation from coal, natural gas, or nuclear produces significant energy losses before reacing electrical components, due to the need to convert heat into mechanical motion. Since ENERGYTRACKR does not account for these losses, its reported energy likely underestimates the actual total.

The tool focuses on short-term test suite execution, which may not capture real-world usage. Nonetheless, measuring energy consumption offers valuable insights for improving software sustainability.

## 7.4. Conclusion validity

We do not make assumptions about the data distribution and instead rely on the nonparametric Wilcoxon rank-sum test and the Vargha-Delaney effect size to quantify the magnitude of differences. An improvement to ENERGYTRACKR would be to assess the normality of measurements directly within the measurement pipeline and automatically schedule non-normal commits for re-measurements. This would ensure a higher number of normally distributed measurements in the final report and allow using stronger statistical tests.

# 8. Discussion

## 8.1. Practical usages for developers

With EnergyTrackr, developers can monitor energy regressions and improvements at the commit level with high precision. The implementation offers a user-friendly report interface and a fully configurable system that supports various projects. Users can measure the energy consumption of their test suite, specific test cases designed to evaluate application behavior, or any command they choose. This analysis can be a one-off or repeated regularly to track the project's energy consumption over time.

Developers can choose between the naive and dichotomous algorithms to analyze their project's evolution. This allows them to balance the need for detailed accuracy with computational and time efficiency; the naive algorithm provides a precise analysis of the software evolution but requires longer processing time, whereas the dichotomous algorithm offers a faster alternative with slightly less precision, while still maintaining reliable confidence in the distributions and trend directions compared to the naive algorithm.

## 8.2. Practical usages for researchers

EnergyTrackr offers several opportunities to better understand how code energy consumption evolves over a commit history. For instance, following a deductive approach, previous work has devoted considerable effort to evaluating how different code patterns influence energy consumption. With EnergyTrackr, we propose to complement this with an inductive approach, traditionally used in repository mining, to identify recurrent patterns in the code change history and its corresponding energy evolution. In our future work, we will explore how commit diffs, which include code, dependency, and configuration changes, can be leveraged to identify new patterns that impact software energy consumption.

## 8.3. Limitations of the current implementation

Using `perf` introduces some limitations, including dependence on RAPL support, which may not be available on all hardware configurations, and the need for specific permissions that are typically restricted in CI environments. Additionally, `perf` is available only on Linux, limiting its applicability to Linux-based systems. A further limitation of the RAPL interface is its inability to provide granular energy consumption data for individual CPU cores. Although there are heuristics for estimating per-core energy usage, these methods were not employed in this analysis because they may reduce overall measurement accuracy relative to the raw RAPL data.

## 8.4. Flaky tests

One *Fastexcel* commit, abc42ae3, was measured with the dichotomous algorithm but not with the naive algorithm, as shown by the orange bar in Figure 4(b). It was excluded from naive measurements due to a test suite failure during pre-measurement build checks. This failure did not occur with the dichotomous algorithm, despite identical server and EnergyTrackr settings, indicating a flaky test (Bell et al., 2018) linked to this commit. Manual log review shows the issue affected 0.43% of commits (1 of 230) with the dichotomous algorithm and was absent in all 1,932 commits tested with the naive algorithm.

We manually reviewed all logs from the measurements (see replication package (Maquoi et al., 2026)) and identified additional flaky tests. Some commits passed initial build checks but showed intermittent failures during measurements. For example, commit a60cc64 from the `Maker` project had 5 failures out of the 30 test suite executions. Overall, 7 of 2,162 commits (0.32%) exhibited such flaky behavior: 2 from `Maker`, 5 from `Fastexcel`. Three `Fastexcel` commits, 7b1525c, and 7b1525c, caused failures across 17 of 64,860 test runs (0.03%). Flaky tests are a known challenge; future work may explore strategies to manage and miticate their impact on measurements.

# 9. Conclusion and Future Work

This paper presents the design, implementation and evaluation of EnergyTrackr, a modular energy measurement and reporting tool based on `perf` and RAPL counters, enabling accurate detection of energy regressions or improvements across project commits. We evaluate two approaches: a naive algorithm that measures and analyzes every commit, and a dichotomous algorithm that samples subset commits. Results show both effectively identify significant energy regressions, with the dichotomous algorithm generally providing statistically and practically comparable results, except for `Simple-OpenAI`. Additionally, the dichotomous approach captures most trend directions identified by the naive mehotd, with minimal omissions.

This work lays the groundwork for future enhancements. EnergyTrackr can be extended to other languages, as it relies on `perf` and commit history, making it language-agnostic. Currently, its granularity is limited to commits, requiring manual inspection to pinpoint energy regressions. However, since `perf` supports high-frequency sampling, it could be adapted to analyze at the function level. Additionally, some flaky tests were identified during measurements. Future versions of EnergyTrackr could incorporate strategies to better handle such tests.

Future work will involve a manual examination of the energy regressions and improvements identified by EnergyTrackr, to uncover code patterns affecting software energy consumption, assessing their impact and generalizability.

# CRediT authorship contribution statement

**Jérôme Maquoi:** Data Curation, Formal Analysis, Investigation, Methodology, Resources, Software, Supervision, Validation, Visualization, Writing – original draft,

Writing – review and editing. **François Bechet:** Conceptualization, Data Curation, Formal Analysis, Investigation, Methodology, Resources, Software, Validation, Visualization, Writing – original draft, . **Luís Cruz:** Conceptualization, Methodology, Resources, Supervision, Validation, Writing – review and editing. **Benoît Vanderose:** Project Administration, Resources, Supervision, Writing – review and editing. **Xavier Devroey:** Funding Acquisition, Methodology, Project Administration, Resources, Supervision, Validation, Visualization, Writing – original draft, Writing – review and editing.

## Declaration of generative AI use

During the preparation of this work, the authors used AI-assisted technologies to polish the language and correct the grammar and LaTeX formatting of the manuscript (goblin.tools and Grammarly). MS Copilot and Lumo were also used in chat mode to help write Python scripts for data analysis and visualization. All content generated with the assistance of these tools was thoroughly reviewed and edited by the authors to ensure it aligned with their intended meaning and contributions. The authors take full responsibility for the content of the published article.

## Data availability

The code, data, and instructions needed to replicate the data analysis are available on Zenodo Maquoi et al. (2026) and GitHub. EnergyTrackr is available on GitHub at https://github.com/snail-unamur/energytrackr/releases/tag/1.0.0.

## Funding sources

This research was supported by the CyberExcellence by DigitalWallonia project (No. 2110186), funded by the Public Service of Wallonia (SPW Recherche).

Jérôme Maquoi is a PhD student at the University of Namur and a member of the SNAIL team, under the supervision of Prof. Benoît Vanderose and Xavier Devroey. His main research interests include green coding, energy consumption, and automated analysis.

François Bechet is a software engineer from the University of Namur. During his master's internship, he worked on green coding as a research trainee at the Delft University of Technology under the supervision of Prof. Luís Cruz.

Luís Cruz is an assistant professor in software engineering at Delft University of Technology and co-founder of GreenSeal.dev, 2600 AA Delft, The Netherlands.

Benoît Vanderose is an associate professor of software engineering at the University of Namur. His main research interests include Agile software development, DevEx, and software quality.

Xavier Devroey is an associate professor of software engineering at the University of Namur. His main research interests include software testing, test suite generation and augmentation, DevOps, and variability-intensive systems.